\documentclass[prd,aps,twocolumn,floatfix]{revtex4-1}
\usepackage{amssymb,graphicx}
\usepackage{epsfig}
\usepackage{amsmath}
\usepackage[usenames]{color}

\begin{document}

\title{ Noncommutative Reissner-Nordstr\o{}m Black hole}

\author{Carlos A. Soto-Campos$^{1}$ and 
Susana Valdez-Alvarado$^{1}$,}
\affiliation{$^{1}${Universidad Aut\'onoma del estado de Hidalgo (UAEH), Hidalgo, M\'exico.\\
\rm{Carretera Pachuca-Tulancingo Km. 4.5, Mineral de la Reforma. C.P. 42184} }
%$^{3}${Centro de Investigaci\'on y de Estudios Avanzados del IPN,  D.F., M\'exico. } \\
}

\begin{abstract}
A deformed embedding of the Reissner-Nordstr\o{}m spacetime is
constructed within the framework of a noncommutative Riemannian
geometry. We find noncommutative corrections to the usual Riemannian expressions for the metric and curvature tensors, which, in the case of the metric, are valid to all  orders in the deformation  parameter. We calculate
the area of the event horizon of the corresponding noncommutative R-N
black-hole, obtaining  corrections up to fourth order in the
deformation parameter for the area of the black-hole. Finally we include some comments on the noncommutative version on one of the second order scalar invariants of the Riemann tensor, the so called  Kretschmann invariant, a quantity regularly used in order to extend gravity to quantum level. 
\end{abstract}

% PACS,
                             % the Physics and Astronomy
\maketitle

%%%%%%%%%%%%%%%%%%%%%%%%%%%%%%%%%%%%%%%%%%%%%%%%%%%%%%%%%%%%
%%%%%%%%%%%%%%%%%%%%%%%%%%%%%%%%%%%%%%%%%%%%%%%%%%%%%%%%%%%%

%-----------------------------------------------------------------------
\section{\label{sec:introduction-1}Introduction}
%----------------------------------------------------------------------
%%%%%%%%%%%%%%%%%%%%%%%%%%%%%%%%%%%%%%%%%%%%%%%%%%%%%%%%%%%%
%%%%%%%%%%%%%%%%%%%%%%%%%%%%%%%%%%%%%%%%%%%%%%%%%%%%%%%%%%%%
In  General Relativity the main assumption is that the space, time
and gravity can be modeled as a single entity, the spacetime. General
Relativity analyzes spacetime as the background in which, electromagnetism, matter and their
mutual influences interact, and it has been used mainly in the study of large
scale phenomena. 

However it is a general  belief that the picture of spacetime as a
pseudo Riemannian manifold $M$ locally modeled as the flat Minkowski
spacetime, should break down at scales of the order of Planck lenght,
$\lambda_p =  {({\hbar}G/c^3)  }^{1/2} \approx 1.6\times 10^{-33} $ cm. A quantum theory of
fields,  attempting to incorporate gravitation, must consider 
limitations on the possible accuracy of localization of  events in
spacetime. A lot of work has been done  on the possible mechanisms
which could lead to such limitations. The  noncommutative geometry has
arised  as an option for a description of quantized spacetime. The
study of noncommutative geometry aquires relevancy in the research of
the quantum nature of spacetime at high energy scales. The idea of
noncommutative spacetime coordinates is old and has been in the
literature from a long time ago \cite{Snyder:1946qz}. A succesful
approach to this topic is provided by the theory of A. Connes, Ref.
\cite{Connes:1994yd} formulated within the framework of $C^*$
algebras. An important advance in the  mathematical framework of
noncommutative geometry of recent years was intrduced by the
deformation quantization of Poisson manifolds by M. Kontsevich
Ref. \cite{Kontsevich:1997vb}.  His work  has led on  to study new
applications of noncommutative geometries to quantum theory. Besides,
Seiberg and Witten Ref.  \cite{Seiberg:1999vs} showed that the
anti-symmetric tensor field arising from massless states of strings
can be described by the noncommutativity of the coordinates of
spacetime 
\begin{equation}
[x^{\mu},x^{\nu}]
 = i \Theta^{\mu \nu} \label{commut}
\end{equation}
where $\Theta^{\mu \nu}$ is a constant antisymmetric tensor. Now the
multiplication of the algebra of functions is given by the Moyal product
\begin{equation}
\displaystyle{(f\star
g)=f(x)exp\big(\frac{i}{2}}\Theta^{\mu
\nu}\overleftarrow{\partial_{\mu}} 
\overrightarrow{\partial_{\nu}}\big)g(x) \,.
\end{equation}
Many recent investigations are oriented towards a formulation of
General Relativity on noncommutative spacetimes. The noncommutative
geometrical approach to gravity could give us insight into a theory of
gravity compatible with quantum mechanics. Much work has been done in
that direction,  Refs.
\cite{Aschieri:2005yw,Aschieri:2005zs,Chamseddine:2000zu,LopezDominguez:2006wd}. 
The noncommutative spacetime with the commutation relation given by
Eq. (\ref{commut}) violates Lorentz symmetry but it was shown  to have
quantum symmetry under the twisted Poincar\'e algebra, see
Ref. \cite{Aschieri:2005zs}.  The abelian twist element 
\begin{equation}
{\cal{F}}= \exp\big(
\frac{-i}{2} \Theta^{\mu \nu} \partial_{\mu}\otimes
\partial_{\nu} \big)
\end{equation}
was used in Refs.\cite{Aschieri:2005zs,Chaichian:2004yh} to twist the
universal enveloping algebra  of the Poincar\'e algebra providing a
noncommutative multiplication for the algebra of functions which is
related to the Moyal product. Then it seems natural to extend this
procedure to other symmetries of noncommutative field theory. 

Related to the noncommutative formulation of General Relativity,
Nicolini et al. Ref. \cite{Nicolini:2005vd}  found a new solution of
the coupled Einstein-Maxwell field equations inspired in the
noncommutative geometry. The metric they have found interpolates
smoothly between a \emph{de Sitter} geometry at short distances, and a
\emph{Reissner-Nordstr\o{}m} (R-N) geometry far away from the
origin. Contrary to the ordinary R-N spacetime, in this metric there
is no curvature singularity at the origin, neither naked nor shielded
by horizons, which seems very intriguing. 

In  a second paper, see Ref. \cite{Ansoldi:2006vg}, the same authors
solved Einstein-Maxwell equations in the presence of a static
spherically symmetric Gaussian distribution of mass and charge having
a minimal width. They show that the coordinate fluctuations can be
described, within the coherent states approach, as a smearing effect. 
%\medskip

In this letter we are going to follow the theory of noncommutative
Riemannian geometry developed by Chaichian et al.  in
Ref. \cite{Chaichian:2007tk} to investigate quantum aspects of gravity
from a mathematical point of view. In Ref. \cite{Chaichian:2007tk}  a
noncommutative Riemannian geometry is constructed by developing the
geometry of noncommutative n-dimensional surfaces. The notions of
metric and conections are introduced on such noncommutative surfaces,
giving rise to the corresponding Riemann curvature. Chaichian et
al. makes use of Nash's theorem of isometric embeddings
---Refs. \cite{Friedman:1961, Nash:1956}---in order to obtain the
deformed versions of classical spacetimes. In the same framework of
noncommutative geometry, Wang et al.  \cite{Wang:2008ut} have
constructed a quantum Schwarzschild spacetime and a quantum
Schwarzschild-de Sitter spacetime  with cosmological constant. They
computed the metrics and curvatures and finally showed that, up to
second order in the deformation parameter, the quantum spacetimes are
solutions of the so called noncommutative Einstein equations. 
%\smallskip
In the present work we will construct  a noncommutative deformation (which can be interpreted as quantum corrections)  of the
Reissner-Nordstr\o{}m spacetime and we will  look for  properties of
such noncommutative spacetime. The key results are the equations of
the metric and the components of the Riemann tensor.  A preliminary version
of this work (exploring a different approach) was presented in the IX Workshop of the Gravitation and
Mathematical Physics Division of the Mexican Physical Society,
\cite{SotoCampos:2011hf}. \\
%\medskip
The paper is organized as follows. In Section \ref{sec:def-alg-funct} we give a quick review
of the deformation of the algebra of functions on the domain of
Euclidean space. In section \ref{sec:def-geom}  we explain how we ``deform'' the
Riemannian geometry by calculating the relevant quantities  in General
Relativity. This particular section  is based  upon Chaichian et al.
\cite{Chaichian:2007tk}. In Section \ref{sec:R-N-sol} the Reisnner-Nordstr\o{}m
spacetime is introduced. We find the components of the noncommutative
versions of the metric and Riemann tensors. We compute the area of the
event horizon and find its noncommutative corrections. Besides, we search for the generalization of the Kretschmann invariant (which provides us information about the embedding) and we find its 
noncommutative expression. Finally in
Section \ref{sec:conc-persp}  we give our conclusions and perspectives.

%%%%%%%%%%%%%%%%%%%%%%%%%%%%%%%%%%%%%%%%%%%%%%%%%%%%%%%%%%%%
%%%%%%%%%%%%%%%%%%%%%%%%%%%%%%%%%%%%%%%%%%%%%%%%%%%%%%%%%%%%

%-----------------------------------------------------------------------
\section{\label{sec:def-alg-funct} Deforming the Algebra of Functions }
%----------------------------------------------------------------------
%%%%%%%%%%%%%%%%%%%%%%%%%%%%%%%%%%%%%%%%%%%%%%%%%%%%%%%%%%%%
%%%%%%%%%%%%%%%%%%%%%%%%%%%%%%%%%%%%%%%%%%%%%%%%%%%%%%%%%%%%
In this section we will introduce the deformation of the
geometry. Keeping this in mind let's begin by introducing a
noncommutative algebra denoted by $\mathcal{A}_{\hbar}$, called
\textit{deformed algebra}, satisfying a {\it correspondence principle}
which stablishes that we recover the  commutative algebra for
$\lim_{{\bar{h}}\rightarrow 0} \mathcal{A}_{\hbar} = \mathcal{A}$. 

 There are different ways to deform a theory. Here we follow the
 approach given by Wang et al. in  Ref. \cite{Wang:2008ut} and we use
 the Riemannian structures previously studied in
 Ref. \cite{Chaichian:2007tk}, using the Moyal product of
 functions. The next subsection is based on the pioneering work of
 Gerstenhaber, Ref. \cite{Gerstenhaber:1963zz}, who  studied the
 deformation of  algebras.

%%%%%%%%%%%%%%%%%%%%%%%%%%%%%%%%%%%%%%%%%%%%%%%%%%%%%%%%%%%%

%-----------------------------------------------------------------------
\subsection{\label{sec:def-alg}Deformation of the Algebra}
%----------------------------------------------------------------------
%%%%%%%%%%%%%%%%%%%%%%%%%%%%%%%%%%%%%%%%%%%%%%%%%%%%%%%%%%%%
Let $\mathcal{A}$ be a commutative ring with the unit. The Ring of
formal power series $R = \mathcal{A}[[{X}]]$ is the set of all the
sequences $(a_{0}, a_{1},a_{2},\dots)$ with $a_{i} \in \mathcal{A}$
for all $i \in \mathbb{N}$ where the operations of sum and product are
defined as: 

1. $(a_{0},a_{1},a_{2},\ldots) + (b_{0},b_{1},b_{2},\ldots)=(a_{0}+b_{0},a_{1}+
b_{1},a_{2}+b_{2},\ldots)$.

2. $(c_{0},c_{1},c_{2},\ldots) = (a_{0},a_{1},a_{2},\ldots)
\times (b_{0},b_{1},b_{2},\ldots)$, where $c_j= a_{0}b_j+a_{1}b_{j-1}+
\ldots + a_jb_{0}  $ for all $j \in \mathbb{N}$. 

It is verified immediately that $R$ is a commutative ring with a unit
$1= (1, 0, 0, \ldots )$. We write $X=(0,1,0,0,\ldots)$, such that
$X^{2}=(0,0,1,0,\ldots) $, etcetera. 

We'll say that $\alpha = (a_{0},a_{1},a_{2},\ldots)$ is of order $i$
when $a_{0}=a_{1}=\ldots = a_{j-1}= 0$ and $a_i\neq 0$. We write
$\cal{O} (\alpha)$ $=  i $. 

We have given a precise definition of a formal power series. From now
on we are going to use the notation $\sum_{i\geq 0} a_i X^{i} $
instead of $(a_{0},a_{1},a_{2},\ldots)$. It is clear that this is not
a sum. Besides,  $\mathcal{A}[[{X}]]$ is a subring of $R$. 

Let's denote by $U$ a certain domain in $\mathbb{R}^{n}$ and  let
$\mathcal{A}$ be the set of all the formal power series in  $\hbar$ 
with coefficients on the real functions  $C^{\infty}$ on $U$. Then 
$\mathcal{A}$ is a  $\mathbb{R}[[\hbar]]-$ module and its elements are
formed as  $\sum_{i\geq 0}f_{i}\hbar^{i}$, where the explicit product
of two elements from the $\mathbb{R}[[\hbar]]-$  module  is:
\begin{equation*}
\left(\sum_{i\geq 0}f_{i}\hbar^{i}\right)\left(\sum_{i\geq
0}g_{i}\hbar^{i}\right)=\sum_{n\in \mathbb{N}}\left(\sum_{k\geq
0}^{n}f_{k}g_{n-k}\hbar^{n}\right)
\end{equation*}
Up to this point we have given the algebra that we are going to use to
quantize (which means: to deform) the spacetime. Now we will proceed
to deform the product of functions substituting the usual product by a
noncommutative one, the so called  \emph{Moyal product} .
%\medskip
Let  $f,g: U \rightarrow \mathbb{R}$ be two functions $C^{\infty}$, we
denote by  $fg$ the usual (i.e. commutative) product of
functions. Then  we  define the star product (or Moyal product) as the
operation $\star:$ $\mathcal{A}\times \mathcal{A}\rightarrow
\mathcal{A}$ such that: 
%\medskip
\begin{equation}
\displaystyle{(f\star
g)=f(x)exp\Big(\hbar
\sum_{i,j}^{n}\Theta_{ij}\overleftarrow{\partial_{i}} 
\overrightarrow{\partial_{j}} \Big) g(x)}
\end{equation}
where the exponencial function
$exp(\hbar \sum_{i,j}^{n}\Theta_{ij}\overleftarrow{\partial_{i}}
\overrightarrow{\partial_{j}})$
must be understand as a power series in the differential operator and
$\Theta_{ij}$ is a constant antysymmetric tensor represented by a
matrix of $(n\times n)$, where $n$ is the dimension of  $U$. It is
easy to see that the star product is associative and satisfies
distributivity too.   ($\mathcal{A}$, $\star$) is the so called
deformed algebra. 

Obviously we can construct more complex structures with the deformed
algebra: let's denote by $\mathcal{A}^{m}=\mathcal{A}\oplus
\mathcal{A}\oplus \cdots \oplus \mathcal{A}$, ($m$ times), with
elements $(X_{1},X_{2},\ldots,X_{m})$, where $X_{i}\in \mathcal{A}$,
with the index $i$ running from 1 to $m$.

%%%%%%%%%%%%%%%%%%%%%%%%%%%%%%%%%%%%%%%%%%%%%%%%%%%%%%%%%%%%
%%%%%%%%%%%%%%%%%%%%%%%%%%%%%%%%%%%%%%%%%%%%%%%%%%%%%%%%%%%%

%-----------------------------------------------------------------------
\section{\label{sec:def-geom}Deformation of the Geometry}
%----------------------------------------------------------------------
%%%%%%%%%%%%%%%%%%%%%%%%%%%%%%%%%%%%%%%%%%%%%%%%%%%%%%%%%%%%
%%%%%%%%%%%%%%%%%%%%%%%%%%%%%%%%%%%%%%%%%%%%%%%%%%%%%%%%%%%%

Now we will construct an embedding  $X$ of a pseudo Riemannian
manifold of dimension  $n$ to a pseudo Euclidean space of dimension
$m$ and with a metric tensor 
\begin{equation}
\eta_{\alpha
\beta}=diag(\underbrace{1,1,\ldots,1}_{p},\underbrace{-1,-1,\ldots,-1}_{q}),
\end{equation}
with $p+q=m$, such that  $X=(X^{1}, X^{2},\ldots
,X^{m})\in\mathcal{A}^{m}$. This section is based entirely in
Ref.\cite{Wang:2008ut}. 
\\
We can construct tangent vectors denoted by $E_{i}$
\begin{equation}
E_{i}\equiv \partial_{i}X,  (i=1, 2, \ldots, n)
\end{equation}
at each point of the surface given by the parametrization $X$.
Now we will define the metric tensor as an $\left(n\times n\right)$ matrix:
\begin{equation}
\mathbf{g}_{ij}= E_{i}\bullet E_{j},\label{metricaNC}
\end{equation}
where the fat dot $\bullet$ denotes \emph{inner product} between
elements of the algebra 
\begin{equation}
\bullet : \mathcal{A}^{m} \otimes \mathcal{A}^{m} \rightarrow
\mathcal{A}^{m}
\end{equation}
on the open region $U$ by
\begin{equation}
A \bullet B= \sum_{i=1}^{p} a_{i}\star b_{i} -\sum_{j=p+1}^{p+q} a_{j}\star b_{j},
\end{equation}
for every  $A=(a_{1},\ldots,a_{m})$ and $B=(b_{1},\ldots,b_{m})$ in
$\mathcal{A}^{m}$.
\\
Obviously  $\mathbf{g}_{ij}$ is invertible in $U$ if and only if
$\mathbf{g}_{ij}|_{\hbar=0}$ is invertible, and we denote the inverse matrix
 $\mathbf{g}^{ij}$, as the matrix which satisfies

$$
\mathbf{g}_{ij}\star \mathbf{g}^{jk}=\mathbf{g}^{kj}\star\mathbf{g}_{ji}=
\delta_{i}^{k},
$$
with  $\delta_{i}^{k}$ the identity $n \times n$ matrix.

%\medskip

%\medskip

In this noncommutative spacetime the relevant connection is given by

$$\nabla_{i}E_{j}= \Gamma_{ij}^{k}\star E_{k},\label{coneccionNC}$$
where the corresponding expression for $\Gamma_{ij}^{k}$ is
\begin{equation}
\mathbf{\Gamma}^k_{ij}= \partial_{i}E_{j}\bullet \tilde{E^k}. \label{gammaNC}
\end{equation}
and $\tilde{E^k} = E_i \star \mathbf{g}^{ik} $. The respective noncommutative Riemann tensor  is given by:
\begin{equation}
\mathbf{R}_{kij}^{l}=
\partial_{i}\mathbf{\Gamma}_{jk}^{l}-\partial_{j}\mathbf{\Gamma}_{ik}^{l}+\mathbf{\Gamma}_{jk}^{p}\star\mathbf{\Gamma}_{ip}^{l}-
\mathbf{\Gamma}_{ik}^{p}\star\mathbf{\Gamma}_{jp}^{l},\label{RiemmanNC}
\end{equation}
In terms of the deformed product, two diferent contractions of the Riemann
tensor can be defined: the \emph{Ricci curvature} tensor
 $\mathbf{R}_{j}^{i}$ and the  $\mathbf{\Theta}^{i}_j$ \emph{curvature tensor}:
\begin{equation}
\mathbf{R}_{j}^{i}=\mathbf{g}^{ik}\star \mathbf{R}_{kpj}^{p},\label{RicciNC}
\end{equation}
\begin{equation}
\mathbf{\Theta}_{p}^{l}=\mathbf{g}^{ik}\star \mathbf{R}_{kpi}^{l}.\label{ThetaNC}
\end{equation}
In general $\mathbf{R}_{j}^{i}$ and $\mathbf{\Theta}_{p}^{l}$ do not coincide, in contrast to the
commutative case.
\\
Now the noncommutative \emph{curvature scalar} of the   surface $X$ is:
\begin{equation}
\mathbf{R}= \mathbf{R}_{i}^{i} \label{NCcurvatscalar}
\end{equation}
On the other hand the noncommutative Einstein equations with cosmological constant on $U$ are:
\begin{equation}
\mathbf{R}_{j}^{i}+
\mathbf{\Theta}_{j}^{i}-\delta_{j}^{i}\mathbf{R}+2\delta_{j}^{i}\Lambda=2\mathbf{T}_{j}^{i}, \label{NCEinsteinEq}
\end{equation}
where $T_{j}^{i}$ is some generalized \emph{energy-momentum} tensor, $\Lambda$
is the cosmological constant. This equation reduces to the Einstein's equation in vaccum
when
$T_{j}^{i}=\Lambda=0$.

%%%%%%%%%%%%%%%%%%%%%%%%%%%%%%%%%%%%%%%%%%%%%%%%%%%%%%%%%%%%

%-----------------------------------------------------------------------
\section{\label{sec:R-N-sol}The Reissner-Nordstr\o{}m Solution}
%----------------------------------------------------------------------
%%%%%%%%%%%%%%%%%%%%%%%%%%%%%%%%%%%%%%%%%%%%%%%%%%%%%%%%%%%%

In the next subsection we give a brief review  of the so called
Reissner-Nordstr\o{}m solution. First we give a sumary of the
structures on the standard commutative case. In order to compare it
with the noncommutative case, we show  the components of the metric
tensor, the Christoffel symbols, Riemann tensor and the scalar
curvature. 
%\medskip

Then in the following  subsection we will find the corresponding noncommutative Riemannian
structures introduced in section 3 for the case of R-N spacetime.

%%%%%%%%%%%%%%%%%%%%%%%%%%%%%%%%%%%%%%%%%%%%%%%%%%%%%%%%%%%%

%-----------------------------------------------------------------------
\subsection{\label{sec:comm-R-N-sptm}Commutative Reissner-Nordstr\o{}m
  Spacetime} 
%----------------------------------------------------------------------  
%%%%%%%%%%%%%%%%%%%%%%%%%%%%%%%%%%%%%%%%%%%%%%%%%%%%%%%%%%%%

The Reissner-Nordstr{\o}m  solution represents the spacetime outside a static,
spherically symmetric charged body carring  an electric charge. It is the unique
spherically symmetric asymptotically flat solution of the Einstein-Maxwell equations.

The metric of the R-N spacetime can be written as:
\begin{equation}
ds^{2}=-\Big(1-\frac{2m}{r}+\frac{e^{2}}{r^{2}}\Big)dt^{2}+\Big(1-\frac{2m}{r}+
\frac{e^{2}}{r^{2}}\Big)^{-1}dr^{2}+r^{2}d \Omega^{2},
\label{RNmetric}
\end{equation}
where $m=\frac{2GM}{c^{2}}$ represents the gravitational mass, with  $G$ the  Newton's
constant and $c$ the speed of light. Here  $e$ stands for the electric charge.

If $e^{2}> m^{2}$ the metric is non-singular everywhere except for the irremovable
singularity at $r= 0 $.
If $e^{2}\leq m^{2}$ the metric has singularities at $r_+ $ and $r_- $ where
$r_{\pm}=m\pm (m^{2}-e^{2})^{\frac{1}{2}}$. For a especially succint review see Ref. \cite{Hawking:1973uf}.
In this letter we will be interested only in the region outside $r_+$.
Here, we will denote the coordinates by: $\mathbf{x_1}=r,\   \mathbf{x_2}=\theta, \ \mathbf{x_3}=\varphi,\  \mathbf{x_4}=t$, (notice the particular choice for the coordinate t). Now, 
the respective Christoffel symbols  $\Gamma_{ij}^{k}=g^{kl}\Gamma_{ijl}$ for this spacetime are given by the well known expressions:
\begin{eqnarray}
\Gamma_{14}^{4} &=&
                    \Gamma_{41}^{4}=-\Gamma_{11}^{1}=\Big(1-\frac{2m}{r}+\frac{e^{2}}{r^{2}}\Big)^{-1} 
\Big(\frac{m}{r^{2}}-\frac{e^{2}}{r^{3}}\Big), \nonumber \\
\Gamma_{22}^{1}  &=& -r\Big(1-\frac{2m}{r}+\frac{e^{2}}{r^{2}}\Big), \nonumber\\  
 \Gamma_{33}^{1}  &=&  -r\sin^{2}\theta\Big(1-\frac{2m}{r}+\frac{e^{2}}{r^{2}}\Big), \nonumber \\
%\Gamma_{33}^{1} & = & -r\sin^{2}\theta(1-\frac{2m}{r}+\frac{e^{2}}{r^{2}}) \nonumber \\
\Gamma_{44}^{1}  &=&  \Big(1-\frac{2m}{r}+\frac{e^{2}}{r^{2}}\Big)
                     \Big(\frac{m}{r^{2}}-\frac{e^{2}}{r^{3}}\Big), 
\nonumber \\  
\Gamma_{12}^{2} &=&  \Gamma_{21}^{2}=\Gamma_{13}^{3}=\Gamma_{31}^{3}=r^{-1},
\nonumber \\
%\Gamma_{12}^{2} & = &
%\Gamma_{21}^{2}=\Gamma_{13}^{3}=\Gamma_{31}^{3}=r^{-1}\nonumber \\ 
\Gamma_{33}^{2} &=& -\sin\theta\cos\theta, \ \ \ \Gamma_{23}^{3}  =
                    \Gamma_{32}^{3}={\cos\theta}/{\sin\theta} \label{Christtof} 
\end{eqnarray}
And the nonzero components of the Riemann tensor are:
\begin{eqnarray}
R_{212}^{1} & = & R_{242}^{4}=-r \Big(\frac{m}{r^{2}}-\frac{e^{2}}{r^{3}} \Big), \nonumber \\
R_{313}^{1} & = & R_{343}^{4}=-r\sin^{2}\theta(\frac{m}{r^{2}}-\frac{e^{2}}{r^{3}}) \nonumber \\
R_{414}^{1} & = &
                  \Big(\frac{-2m}{r^{3}}+\frac{3e^{2}}{r^{4}}\Big)\Big(1-\frac{2m}{r}+\frac{e^{2}}{r^{2}}\Big),
                  \nonumber \\    
R_{323}^{2}  & = & \sin^{2}\theta-\sin^{2}\theta \Big(1-\frac{2m}{r}+\frac{e^{2}}{r^{2}}\Big), 
\nonumber \\
R_{121}^{2} & = & -r^{-1}\Big(\frac{m}{r^{2}}-\frac{e^{2}}{r^{3}}\Big) \Big(1-\frac{2m}{r}+\frac{e^{2}}
{r^{2}}\Big)^{-1},  \nonumber \\
%R_{323}^{2}& = & \sin^{2}\theta-\sin^{2}\theta(1-\frac{2m}{r}+\frac{e^{2}}{r^{2}}) \nonumber \\
R_{424}^{2} & = & R_{434}^{3}=r^{-1}\Big(\frac{m}{r^{2}}-\frac{e^{2}}{r^{3}}\Big) \Big(1-
\frac{2m}{r}+\frac{e^{2}}{r^{2}}\Big), \nonumber \\ 
R_{131}^{3}&=& r^{-1}\Big(\frac{m}{r^{2}}-\frac{e^{2}}{r^{3}}\Big) \Big(1-
\frac{2m}{r}+\frac{e^{2}}{r^{2}}\Big)^{-1}\, , \nonumber \\
R_{232}^{3}  &=& 1-\Big(1-\frac{2m}{r}+\frac{e^{2}}{r^{2}}\Big), \nonumber \\ 
 R_{141}^{4} & = & -\Big(\frac{-2m}{r^{3}}+\frac{3e^{2}}{r^{4}}\Big) \Big(1-\frac{2m}{r}+\frac{e^{2}}
{r^{2}}\Big)^{-1},  \label{Riemann}
%R_{141}^{4} & = & -(-\frac{-2m}{r^{3}}+\frac{3e^{2}}{r^{4}})(1-\frac{2m}{r}+\frac{e^{2}}{r^{2}})^{-1}
\end{eqnarray}
\noindent whereas the nonzero components of the Ricci tensor are:
%\medskip
$R_{11}=-\frac{e^{2}}{r^{4}}(1-\frac{2m}{r}+\frac{e^{2}}{r^{2}})^{-1},
\ \ R_{22}=\frac{e^{2}}{r^{2}}, \ \
R_{33}=\frac{e^{2}}{r^{2}}\sin^{2}\theta, \ \
R_{44}=\frac{e^{2}}{r^{4}}(1-\frac{2m}{r}+\frac{e^{2}}{r^{2}})$ 
%\medskip
\noindent And the scalar curvature is  $R=0$.

%%%%%%%%%%%%%%%%%%%%%%%%%%%%%%%%%%%%%%%%%%%%%%%%%%%%%%%%%%%%
%%%%%%%%%%%%%%%%%%%%%%%%%%%%%%%%%%%%%%%%%%%%%%%%%%%%%%%%%%%%

%-----------------------------------------------------------------------
\subsection{\label{sec:noncomm-R-N-sptm}Noncommutative
  Reissner-Nordstr\o{}m Spacetime} 
%----------------------------------------------------------------------
%%%%%%%%%%%%%%%%%%%%%%%%%%%%%%%%%%%%%%%%%%%%%%%%%%%%%%%%%%%%
%%%%%%%%%%%%%%%%%%%%%%%%%%%%%%%%%%%%%%%%%%%%%%%%%%%%%%%%%%%%
  
We are going  to compute the deformed  Riemannian structures given in
section \ref{sec:def-geom} for the R-N spacetime.  By following the procedure depicted
in the preceding section, we propose that the Reissner-Nordstr\o{}m
spacetime can be embedded in a six dimensional pseudo Euclidean space
with metric  $\eta_{\mu\nu}= diag(1,1,1,1,-1,-1)$, through the map
$X=(X_{1},X_{2},X_{3},X_{4},X_{5},X_{6})$ given by 
\begin{eqnarray}
X_{1} & = & g(r) \nonumber \\
X_{2} &  = & r\sin\theta\cos\varphi \nonumber \\
X_{3} & = & r\sin\theta\sin\varphi \nonumber \\
X_{4} & = & r\cos\theta \nonumber \\
X_{5} & = & \Big(1-\frac{2m}{r}+\frac{e^{2}}{r^{2}}\Big)^{\frac{1}{2}}\sin t \nonumber \\
X_{6} & = & \Big(1-\frac{2m}{r}+\frac{e^{2}}{r^{2}}\Big)^{\frac{1}{2}}\cos t \label{X's}
\end{eqnarray}
with the convention of coordinates given in the previous subsection. Here we have used the Kasner embedding already used before by Wang and Zhang in Ref.\cite{Wang:2008ut} for the case of a noncommutative  embedding of a Schwarzschild and a Schwarzschild-de Sitter spacetime in a 6 dimensional pseudo Euclidean manifold. Using a similar
procedure than that implemented by Wang and Zhang, we define the function $g(r)$ in the
first of the Eqs.(\ref{X's}) as a smooth function such that
$$g'^{2}+1=\Big(1-\frac{2m}{r}+\frac{e^{2}}{r^{2}}\Big)^{-1}\Big(1+\Big(\frac{m}{r^{2}}-\frac{e^{2}}{r^{3}}\Big)^{2}\Big)$$ 
where $g'$ denotes the derivative of $g$ with respect to $r$. This requirement for $g(r)$ simplifies the computations of the components of the noncommutative metric as will be clear very soon.
From the Eqs. (\ref{RNmetric}) and (\ref{X's}) it is easy to verify
the isometry of the embedding: 
\begin{eqnarray}
ds^{2} & = & (dX_{1})^{2}+(dX_{2})^{2}+(dX_{3})^{2}+(dX_{4})^{2}  \nonumber \\ 
 {} &{} & -(dX_{5})^{2}-(dX_{6})^{2}
\end{eqnarray}
We now proceed to deform the algebra of functions with the  procedure
mentioned in section  \ref{sec:def-geom} and imposing a Moyal product of functions with
the  matrix representation of the antisymmetric tensor  $\Theta_{\mu\nu}$
\begin{eqnarray}
\Theta_{\mu\nu}=\left(
  \begin{array}{cccc}
    0 & 0 & 0 & 0 \\
    0 & 0 & 1 & 0 \\
    0 & -1 & 0 & 0 \\
    0 & 0 & 0 & 0 \\
  \end{array}
\right) \label{Theta}
\end{eqnarray}
The tangent vectors $E_{i}=\partial_{i}X=\frac{\partial}{\partial x_i}X$ are given by:

\begin{eqnarray}
 E_{1} & = & \Big[
             g'(r),\sin\theta\cos\varphi,\sin\theta\sin\varphi,\cos\theta,\Big(1-\frac{2m}{r}+
             \frac{e^{2}}{r^{2}}\Big)^{-\frac{1}{2}}  \nonumber \\   
    {} &{} & \times \Big(\frac{m}{r^2}-\frac{e^{2}}{r^{3}}\Big)\sin t, \Big(1-\frac{2m}{r}+\frac{e^{2}}{r^{2}}\Big)^{-\frac{1}{2}}\Big(\frac{m}{r^2}- \frac{e^{2}}{r^{3}}\Big)\cos t\Big] \nonumber \\
E_{2} & = & \big[0,r\cos\theta\cos\varphi,r\cos\theta\sin\varphi,-r\sin\theta,0,0\big] \nonumber \\
E_{3} & = & \big[0,-r\sin\theta\sin\varphi,r\sin\theta\cos\varphi,0,0,0\big] \nonumber \\
E_{4} & = & \Big[0,0,0,0,\Big(1-\frac{2m}{r}+\frac{e^{2}}{r^{2}}\Big)^{\frac{1}{2}}\cos t,-\Big(1-\frac{2m}{r}+
\frac{e^{2}}{r^{2}}\Big)^{\frac{1}{2}}\sin t\Big] \nonumber \\ 
\end{eqnarray}

And from Eq.  (\ref{metricaNC}) we find Ref.\cite{SotoCampos:2011hf}     the nonzero components of the metric tensor : 
\begin{eqnarray}
\mathbf{g}_{11} & = & \Big(1-\frac{2m}{r}+\frac{e^{2}}{r^{2}}\Big)^{-1}-\cos{2\theta} \sinh^2\hbar \nonumber \\
\mathbf{g}_{12} & = & \mathbf{g}_{21}= r\sin 2 \theta \sinh^{2}\hbar \nonumber \\
\mathbf{g}_{13} & = & -\mathbf{g}_{31}= - r\sin 2\theta \sinh \hbar\cosh \hbar \nonumber \\
\mathbf{g}_{22} & = & r^{2}+r^{2}\cos 2\theta\sinh^{2}\hbar \nonumber \\
\mathbf{g}_{23} & = &-\mathbf{g}_{32}=-r^{2}\cos 2\theta\sinh\hbar\cosh\hbar \nonumber \\
\mathbf{g}_{33} & = & r^{2}\sin^{2}\theta-r^{2}\cos 2\theta\sinh^{2}\hbar \nonumber \\
\mathbf{g}_{44} & = & -\Big(1-\frac{2m}{r}+\frac{e^{2}}{r^{2}}\Big) \label{metricNC}
\end{eqnarray}
In agreement with the observation of Chaichian et
al. Ref.\cite{Chaichian:2007we}, in the sense that, for an arbitrary
$\Theta_{\mu\nu}$ given by Eq.(\ref{Theta}), the deformed metric
$\mathbf{g}_{\mu \nu}$ is not diagonal.  
In the components of $\mathbf{g}_{\mu \nu}$ we notice that terms
containing the noncommutative parameter appear. Actually the
hyperbolic functions ---depending on the noncommutative parameter
$\hbar $--- enter when we compute the deformed inner product,
Eq. (\ref{metricaNC}), in the form of series expansions.  It is a
straightforward calculation to verify that the components of  the
noncommutative metric tensor satisfy the correspondence principle. It
is important to remark that the components $\mathbf{g}_{12}, \dots ,
\mathbf{g}_{33}$ coincide with those computed by Wang et al \cite{Wang:2009ju}. That
coincidence in some of the components of the metric tensor is due to
the map defined by Eq. (\ref{X's}).     
\smallskip
In order to make these calculations more clear, we now proceed to compute
in detail one of the components of the metric, for example
$\mathbf{g}_{11}$. From Eq.(\ref{metricaNC}) we know that 
$$\mathbf{g}_{11}= E_{1}\bullet E_{1},$$
and using the expressions for the generators obtained before, then  after some simplifications of terms, we get 
\begin{widetext}
\begin{eqnarray}
\mathbf{g}_{11} & & = g'^{2}(r)+\sin^{2}\theta\cos^{2}\varphi+\sin^{2}\theta\sin^{2}\varphi +
  2\frac{\hbar^{2}}{2!}\big[(\sin^{2}\theta\cos^{2}\varphi-\cos^{2}\theta\sin^{2}\varphi)  
+ (\sin^{2} \theta \sin^{2}\varphi-\cos^{2}\theta\cos^{2}\varphi)\big] + \nonumber \\
&+& 8 \frac{\hbar^{4}}{4!}\big[ (\sin^{2}\theta\cos^{2}\varphi-\cos^{2}\theta\sin^{2}\varphi)
+ (\sin^{2} \theta \sin^{2}\varphi-\cos^{2}\theta\cos^{2}\varphi)\big] +\ldots+\cos^{2}\theta -\Big(1-\frac{2m}{r}+\frac{e^{2}}{r^{2}}\Big)^{-1}\Big(\frac{m}{r^{2}}-\frac{e^{2}}{r^{3}}\Big)^{2} \nonumber \\
&{}& 
%32(\sin^{2} \theta \sin^{2}\varphi-\cos^{2}\theta\cos^{2}\varphi)] +\ldots+\cos^{2}\theta %-\Big(1-\frac{2m}{r}+\frac{e^{2}}{r^{2}}\Big)^{-1} \Big(\frac{m}{r^{2}}-\frac{e^{2}}{r^{3}}\Big)^{2}
\end{eqnarray}
\end{widetext}
Now using the fact that
$$g'^{2}+1= \Big(1-\frac{2m}{r}+\frac{e^2}{r^2}\Big)^{-1}\Big(1+
\Big(\frac{m}{r^2}-\frac{e^2}{r^3}\Big)^{2}\Big)$$
and identifying the series  $\sinh^2\hbar= 2\frac{\hbar^2}{2!}+
8\frac{\hbar^4}{4!}+32\frac{\hbar^6}{6!}+\ldots $; we finally obtain the component
$\mathbf{g}_{11}=(1-\frac{2m}{r}+\frac{e^2}{r^2})^{-1}-\cos2\theta \sinh^2\hbar$. Now from Eq.  (\ref{metricaNC}) and using the result $\mathbf{g}^{kj}\star\mathbf{g}_{ji}=
\delta_{i}^{k},$    we  can calculate the  contravariant components of the noncommutative metric tensor now. Those components turn a little bit more complicated than those in Eq.(\ref{metricNC}), precisely because their expressions  involve the determinant of the noncommutative metric. Fortunately the components of the covariant metric tensor  $ \mathbf{g}_{\mu\nu}$ just involve functions of the coordinates $r$ and  $\theta$ and then, the Moyal products between them turn into the usual commutative products. 
%so  det $\mathbf{g}_{\mu\nu} = r^4 \sin^2\theta [P^{-1}(r)
%(1-\cos{2\theta}\sinh^{2}\hbar)+\sinh^{2}\hbar(\cosh^2\hbar +
%2\cos^2\theta)]$,
In the following computations it will be very convenient to introduce the function $P(r)$ which is the metric coefficient of the timelike coordinate in the line element (Eqn.\ref{RNmetric})  defined by $P (r) \equiv
(1-\frac{2m}{r}+\frac{e^{2}}{r^{2}})$. Now we can see that all the components of the noncommutative contravariant metric tensor will be affected by the presence of the charge in the R-N black hole, because of its dependence on the determinant of the noncommutative metric and on the function $P(r)$ defined before.

Now, from  Eq. (\ref{gammaNC}) we can calculate the respective
 components of the noncommutative connection
 $\Gamma^k_{ij}= \partial_{i}E_{j}\bullet \tilde{E^k}$. Taking into account the next definitions: $ P'(r) = \frac{dP}{dr} =
\frac{2m}{r^2} -\frac{2\epsilon^2}{r^3}$, and  $ P''(r) =
-\frac{4m}{r^3} +\frac{6\epsilon^2}{r^4}\, $, the modified connections  will be written  in the form $\mathbf{\Gamma}^\alpha_{\mu\nu} = \Gamma^\alpha_{\mu\nu} + \hbar f_1(x^\mu) + {\hbar}^2 f_2(x^\mu) + ...  $, up to second order in the deformation parameter by : 
\begin{eqnarray}
\mathbf{\Gamma}^1_{11} &=& \Gamma^1_{11} + \frac{P^{'}}{2}(P^{-1}\cos2\theta+1+2\cos^2\theta)\hbar^2, \nonumber \\
\mathbf{\Gamma}^1_{22} &=& \Gamma^1_{22} +\big[ \ rP ( P^{-1}\cos2\theta 
+ 1+  2\cos^2\theta ) \nonumber \\ 
              &+& 2 + 4\cos2\theta \big] \hbar^2, \nonumber \\  
\mathbf{\Gamma}^1_{33} &=& \Gamma^1_{33} + rP \big[ \ P \sin^2\theta ( P^{-1}\cos2\theta 
+ 1+  2\cos^2\theta ) \nonumber \\ 
              &+& 2 (\cos^4\theta +1) + 4\cos2\theta \big] \hbar^2, \nonumber \\
\mathbf{\Gamma}^1_{44} &=& \Gamma^1_{44} +\frac{PP'}{2} \big( P^{-1}\cos2\theta 
+ 1+  2\cos^2\theta \big) \hbar + \mathcal{O}(\hbar^2)  \nonumber \\
                 % &+&1+2 \cos^2\theta \big)
                  %\hbar^2, \nonumber \\
 % \mathbf{\Gamma}^2_{11} &=& \cot\theta\Big(\frac{P'P^{-1}}{r} \cos^2\theta+\frac{2cos2\theta}{r^2}\Big) \hbar^2,  \nonumber \\ 
\mathbf{\Gamma}^2_{12} &=& \Gamma^2_{12} + \frac{1}{r} \big( \csc^2\theta\cos2\theta + 2\cot\theta \cos2\theta  \big) \hbar + \mathcal{O}(\hbar^2) , \nonumber \\
%\mathbf{\Gamma}^2_{22} &=& 2\cot\theta \cos^2\theta(P-1) \hbar^2,
    %             \nonumber \
%\end{eqnarray}
%\begin{eqnarray}
%\mathbf{\Gamma}^2_{23} &=& - \cot\theta \ \hbar
 %                 -\cot\theta
  %                \left(\frac{2}{3} + \cos2\theta
   %               -2P\cos^2\theta \right)\hbar^3 , \nonumber \\ 
\mathbf{\Gamma}^2_{33} &=& \Gamma^2_{33} +
                  \Big[ 2P\cos^3\theta \sin\theta- P\cos\theta \big(-P^{-1}\cos2\theta \cos^2\theta  \nonumber \\ 
                &+& 2\sin^2\theta \big)  + 4\cot\theta\cos2\theta + 2\sin2\theta  \Big] \hbar^2 , \nonumber \\ 
% \mathbf{\Gamma}^2_{44} &=& -\frac{P'P}{r}\cot\theta \cos^2\theta \hbar^2
 %                 \nonumber \\
% \mathbf{\Gamma}^3_{11} &=& -\frac{P^{-1 }P'}{r}\cot\theta \ \hbar -
 %                 \frac{\cot\theta}{r} \Big[
  %                \frac{P^{-1}P'}{3} \nonumber \\
   %              &-&  P'(-P^{-1}\cos2\theta + 1+2 \cos^2\theta) \Big]\hbar^3 ,
    %              \nonumber \\ 
           \mathbf{\Gamma}^3_{13} &=& \Gamma^3_{13} -\frac{P}{r} \big(\cos2\theta+1 \big)\hbar + \Big[-3P(1+ cos^2\theta)+\cos2\theta    \nonumber \\
           & \times & \big(3 - \cot^2\theta + 2\cot\theta \big)  -  2\cot\theta \big(\sin2\theta+2\cos\theta \big)\Big]\hbar^2     , \nonumber \\
% \mathbf{\Gamma}^3_{22} &=& - 2(P-1)\cot\theta \ \hbar -
 %                2\cot\theta
  %                \Big[\frac{2}{3}(P-1) \nonumber \\
   %               &+& (1-P)P \big(-P^{-1}\cos2\theta +
    %              1 +2 \cos^2\theta\big)\Big]\hbar^3 , \nonumber \\
  \mathbf{\Gamma}^3_{23} &=& {\Gamma}^3_{23} + P\cot\theta \big(\cos2\theta+1 \big)\hbar  + \Big[ \cot\theta \Big(-3P
                  \nonumber \\
                  &-&   2 \cot\theta\sin2\theta  +3\cos2\theta - 2P\cos^2\theta-7 \cot^2\theta              \nonumber \
% \mathbf{\Gamma}^3_{33} &=& \Big[\cot\theta+
 %                 2\cos\theta\sin\theta(1-P)\Big]\hbar \nonumber \\
% &+&\Big[         \cot\theta
  %                \Big((P+\frac{2}{3}) -P (-P^{-1}\cos2\theta  + 1
  %                \nonumber \\
   %           &+&2\cos^2\theta) \Big)+2 \cos\theta \sin\theta 
    %              \Big(\frac{2}{3}(1-P) \nonumber \\
      %            &-& (1-P)P (-P^{-1}\cos2\theta
      %            + 1+2\cos^2\theta) \Big) \Big]\hbar^3 ,  \nonumber \\ 
 % \mathbf{\Gamma}^3_{44} &=& \frac{PP'}{r}\cot\theta \ \hbar
   %                +\Big[\frac{2}{3}\frac{PP'}{r}\cot\theta
    %               -\frac{P^2P'}{r}\cot\theta \nonumber \\
     %            &\times&  (-P^{-1}\cos2\theta
      %            + 1 +2 \cos^2\theta \Big] \hbar^3, \nonumber \\
      \end{eqnarray}
  
      \begin{equation}
      + \  2 \frac{\cos(2\theta)}{\sin\theta} +\frac{2}{\sin^2\theta} -2\Big) + 4P\cos\theta\sin\theta\Big]\hbar^2 \hspace{1.6cm} \nonumber  \
      \end{equation}
       \noindent and finally,  
      \begin{eqnarray} 
    \mathbf{\Gamma}^4_{14} &=& {\Gamma}^4_{14}. \hspace*{7.0cm} \label{ChristtofNC}
\end{eqnarray} 

%\end{widetext}
\noindent 
%where the function $D$ is defined as $D\equiv  P^{-1} %(1-\cos{2\theta}\sinh^{2}\hbar)+\sinh^{2}\hbar(\cosh^2\hbar + %2\cos^2\theta)$. 
where we have used the fact that, by definition, $\mathbf{\Gamma}^k_{ij}$ is symmetric in the covariant indices. This last statement is clear from Eq. (\ref{gammaNC}).    It is interesting to
compare the expressions that we have just  obtained in Eq. (\ref{ChristtofNC}) with those, previously listed in Eq. (\ref{Christtof}). We can easily  verify that the correspondence principle is satisfied. With the last expressions
given in Eq. (\ref{ChristtofNC}) we calculate the components of the
noncommutative Riemann tensor. 

In order to compare with the
respective commutative expressions ---Eq.(\ref{Riemann}) in subsection
\ref{sec:comm-R-N-sptm}--- we focus only those components. However there are more components of the Riemann tensor which are purely noncommutative and don't have a commutative counterpart. 
For the sake of clarity in the expressions we will write one of the components of noncommutative Riemann tensor in the form $\mathbf{R}^\alpha_{\beta \mu\nu} = R^\alpha_{\beta \mu\nu} + \hbar f_1(x^\mu) + {\hbar}^2 f_2(x^\mu) + ...  $

\begin{eqnarray}
\mathbf{R}^t_{rtr} &=&    - \partial_r \mathbf{\Gamma}^t_{rt} + \mathbf{\Gamma}^r_{rr} \star \mathbf{\Gamma}^t_{rt} - \mathbf{\Gamma}^t_{rt} \star \mathbf{\Gamma}^t_{rt} \nonumber \\ 
&=&  {R}^t_{rtr}+\frac{P^{-1}P^{'2}}{4}(1+2\cos^2\theta) \hbar^2  \label{Riem4141}
\end{eqnarray} 
\noindent  As in the case of the noncommutative conections, we can verify that the
correspondence principle is satisfied for the limit
${\hbar}\rightarrow 0$. 
For the rest of the noncommutative Riemann tensors, the calculations are straightforward but the expressions rapidly become cumbersome in some cases. For many of the noncommutative Riemann tensors, the first nonvanishing corrections appear at second order in $\hbar$. However, some of the purely noncommutative Riemann tensors have expressions which depend on first order terms of $\hbar$. 

One useful quantity that we can calculate is the Kretschmann scalar $K = \mathbf{R}^{\alpha\beta\mu\nu} \star \mathbf{R}_{\alpha\beta\mu\nu}$.  This invariant has been used in the pursue to extend gravity to quantum level.

 In our case there are some of the noncommutative components of the totally covariant Riemann tensor that contribute to first order in the perturbative parameter $\hbar$. For example, those depending on the connections $\mathbf{\Gamma}^1_{44} $, $\mathbf{\Gamma}^2_{12} $, $\mathbf{\Gamma}^3_{13} $ and $\mathbf{\Gamma}^3_{23} $  written in Eq.(\ref{ChristtofNC}). Thus we expect corrections of order $\hbar$ in $\mathbf{R}_{1 \beta\mu\nu}$, $\mathbf{R}_{2  \beta\mu\nu}$ and $\mathbf{R}_{3 \beta\mu\nu}$. So first order corrections to the Kretschmann scalar come form cross products of the usual contravariant Riemann components with their respective covariant first order corrections and vice versa. 
 
 After performing this long calculation we find that 
 
 $$K_{NC} = K + f(r,\theta,\phi) \hbar $$
 where $f(r, \theta, \phi)$ comprises the first order corrections of all the noncommutative Riemann tensors and    $K$ is the usual commutative expression reported in ( \cite{Cherubini:2003nj, 0004-637X-535-1-350} )

 \medskip
 At this point we can obtain various
expressions for different physical quantities in terms of the
deformation parameter. Although the Hawking temperature of the R-N
black hole is not modified by the embedding, we can ask us about the
area of the event horizon in the deformed R-N Black Hole. This is
given by the integral 
$$
A= \int_{r=r_+} \sqrt{ \det{\mathbf{g}_{ab}}} \ d\theta  d\phi
$$
with $a$, $b$ $=2, 3$, and $r_+$ stands for the exterior radius of the event horizon $r_+= m + \sqrt{m^{2}-e^{2}}$.  From Eq.(\ref{metricNC}), we write
\begin{eqnarray}
  \mathbf{g}_{ab}=\left(
  \begin{array}{cccc}
    \mathbf{g}_{22} & \mathbf{g}_{23}  \\
    \mathbf{g}_{32} & \mathbf{g}_{33}  \\
  \end{array}
  \right)
\end{eqnarray}
Which leads to the following result 
\begin{eqnarray}
  A  =   \int_{r=r_+} r^2 \sin{\theta}  \sqrt{ 1-\cos{2\theta}\sinh^2{\hbar} } \ d\theta  d\phi ,
\end{eqnarray}
Performing the integral in $\theta$ we obtain the following expression
in terms of the parameter $\hbar$  
\begin{eqnarray}
  \sqrt{(1-\sinh^2{\hbar})} & +& \frac{\sqrt{2}}{2\sinh{\hbar}} \arctan{\Big(\frac{\sqrt{2} \sinh{\hbar}}{\sqrt{1-\sinh^2{\hbar}}}\Big)}   \nonumber \\ 
  {} & + &\frac{\sqrt{2}}{2}\sinh{\hbar} \arctan{\Big(\frac{\sqrt{2} \sinh{\hbar}}{\sqrt{1-\sinh^2{\hbar}}}\Big)} \nonumber
\end{eqnarray}
Which is a smooth function of $\hbar$. In order to compare with
the usual commutative result, this can be expanded in powers of the
deformation parameter using the series expansion  of $\sinh^2{\hbar}$ given before and the corresponding expression for  the function $\arctan(x)$. Then, for the area of the event horizon of the deformed R-N
black hole we obtain  
 \begin{eqnarray}
 %A & = &  \int_{r=r_+} r^2 \sin{\theta}  \sqrt{ 1-\cos{2\theta}\sinh^2{\hbar} } \ d\theta  d\phi \nonumber \\
 A  & = &  4\pi r_+^2  \Big(1  + \frac{\hbar^2 }{6} - \frac{ \hbar^4}{360} +   \mathcal{O}(\hbar^6) \Big)
  \end{eqnarray}
This result is similar to that obtained by Wang and Zhang in
Ref.\cite{Wang:2008ut}. After all this is not surprising, because the determinant
of $\mathbf{g}_{ab}$ has the same expression  as that reported in  \cite{Wang:2008ut}. That is due
to the particular embedding we have used which inherits the  symmetry of Schwarzschild solution over the angular coordinates. It is easy to verify that we obtain the regular ---commutative--- expression for the case in which ${\hbar}\rightarrow 0$. 
However,  in the present  case, the radius of the event horizon is obviously modified by the charge of the
black hole. 
Then, for the  expressions of the components of the
Riemann tensor ---see for example Eq.(\ref{Riem4141})---  we have nontrivial modifications because of the
presence of the functions $P(r)$ and its derivatives, which contain explicitly the charge. In the same way we expect nontrivial modifications for the
deformed Einstein field equations Eq.(\ref{NCEinsteinEq}).

%%%%%%%%%%%%%%%%%%%%%%%%%%%%%%%%%%%%%%%%%%%%%%%%%%%%%%%%%%%%

%-----------------------------------------------------------------------
\section{\label{sec:conc-persp}Conclusions and perspectives}
%----------------------------------------------------------------------
%%%%%%%%%%%%%%%%%%%%%%%%%%%%%%%%%%%%%%%%%%%%%%%%%%%%%%%%%%%%
As we have mentioned before, there are different ways to construct a
noncommutative geometrical approach to gravity. In this letter we have
followed the prescription given by Wang and Zhang in
Ref.\cite{Wang:2008ut}. We don't find corrections to the surface gravity of the noncommutative RN black hole:  $\kappa = 1/2 \partial_1 \mathbf{g}_{44} |_{r = r_+}$, so the Hawking temperature

$$T_H = \frac{ \sqrt{m^2 - e^2} }{2\pi(m+\sqrt{m^2 - e^2})^2 } $$ 
is not modified by the embedding, in contrast to Ref. \cite{Ansoldi:2006vg} where a
correction to $T_H$ is given in terms of the noncommutative parameter
for a variety of charged objects. On the other hand, we have found an
analytic expression for the noncommutative corrections to the event
horizon area, related to the entropy of the R-N black hole. We
developed the calculations of the noncommutative Ricci tensors, but
the mathematical expressions that we obtain become very heavy quite soon.

For the case of the
deformed versions of Einstein field equations,
Eq.(\ref{NCEinsteinEq}), it is not clear how to construct a suitable
noncommutative stress tensor $T^i_j$. A possible solution to this
problem would be to proceed like in Ref. \cite{Wang:2008ut}, working
with approximate expressions ---given in terms of powers of the noncommutative parameter, in the same way that we have proceeded in this letter---  for the components of the Ricci
tensor. This will be studied in a future paper.  

%{\bf{Include comments on Bertolami \& Guisado!!} }

%{\bf{Include comments on Alfredo's paper!!} }
%%%%%%%%%%%%%%%%%%%%%%%%%%%%%%%%%%%%%%%%%%%%%%%%%%%%%%%%%%%%
%%%%%%%%%%%%%%%%%%%%%%%%%%%%%%%%%%%%%%%%%%%%%%%%%%%%%%%%%%%%

%-----------------------------------------------------------------------
\section{\label{sec:acknwldgmnts}Acknowledgements}
%----------------------------------------------------------------------
We would like to thank to Hugo Garc\'ia Compe\'an for very helpful suggestions and for pointing us some of the references of this paper. We also thank to Luis Lopez for helpful discussions. 

The work of C. S.  was partially supported by Prodep. S. V.  was supported by a Posdoctoral Grant of Prodep, MEXICO. 
%%%%%%%%%%%%%%%%%%%%%%%%%%%%%%%%%%%%%%%%%%%%%%%%%%%%%%%%%%%%

%-----------------------------------------------------------------------
\
%%%%%%%%%%%%%%%%%%%%%%%%%%%%%%%%%%%%%%%%%%%%%%%%%%%%%%%%%%%%
\bibliographystyle{aa}
\bibliography{refart}% Produces the bibliography via BibTeX.
%%%%%%%%%%%%%%%%%%%%%%%%%%%%%%%%%%%%%%%%%%%%%%%%%%%%%%%%%%%%

\end{document}